\documentstyle[12pt,epsfig]{article}
\begin{document}
\oddsidemargin 0mm
\evensidemargin 0mm
\voffset-20mm
\hoffset10mm
\pagestyle{plain}
\begin{center}\Large\bf
%-----------------------------------------------------------------
%  Title
%-----------------------------------------------------------------
Invisible Clusters and CMB Decrements
%-----------------------------------------------------------------
\end{center}
\begin{center}
%-----------------------------------------------------------------
%  Authors
%-----------------------------------------------------------------
Youri Dabrowski
%-----------------------------------------------------------------
\end{center}

\begin{center}
%-----------------------------------------------------------------
%  Institutes
%-----------------------------------------------------------------
Mullard Radio Astronomy Observatory, Cavendish Laboratory,
Madingley Road, Cambridge, CB3 0HE, UK
%-----------------------------------------------------------------
\end{center}
%-----------------------------------------------------------------
%  Abstract here: maximum 200 words
%-----------------------------------------------------------------

\begin{abstract}
A decrement in the Cosmic Microwave Background (CMB) 
has been observed by the Ryle Telescope towards
a pair of, possibly lensed, quasars (PC1643+4631 A\&B).
Assuming that the decrement is due to the Sunyaev-Zel'dovich (S-Z)
effect, this is indicative of a very rich intervening cluster,
although no X-ray emission has yet been observed in that direction.
In order to investigate these problems, we present a new model
for the formation of distant spherically symmetric clusters in an
expanding Universe. Computation of photon
paths allows us to evaluate the gravitational effects on
CMB photons passing through the evolving mass (i.e. Rees Sciama
effect).
The lensing properties of the cluster are also considered so that
the model can be applied to the PC1643+4631 case to retrieve
both the S-Z flux and the separation of the quasar pair.
We find that the Rees Sciama effect might contribute significantly
to the overall observed CMB decrement.
\end{abstract}

%-----------------------------------------------------------------
%  text
%-----------------------------------------------------------------

%-----------------------------------------------------------------
%  end of abstract
%-----------------------------------------------------------------

\section{Introduction}
In this paper we model the formation of high redshift galaxy
clusters in order to investigate their effect on the CMB.
Together with the S-Z effect, we will be concerned with another
secondary anisotropy which was first discussed by
Rees \& Sciama (1968). These authors pointed out that there is
a gravitational effect on the CMB photons while they are crossing
evolving cosmic structures such as a collapsing cluster of galaxies,
since photons climb out of a slightly different potential well
than the one into which they entered.
In the next section, we present the observations carried out
by Jones et al. (1997) which suggest the presence of a rich cluster
of galaxies at a redshift $z\geq 1$.
Section \ref{model} presents an improved model for the
formation of spherical galaxy clusters. This model is applied
to the PC1643+4631 observations and the results are summarised
in section \ref{results}.
The work overviewed in this paper is discussed in
more details in forthcoming papers (Lasenby et al. 1997
\& Dabrowski et al. 1997).

\section{Observations}
\label{observations}
\begin{figure}
\centerline{\epsfig{file=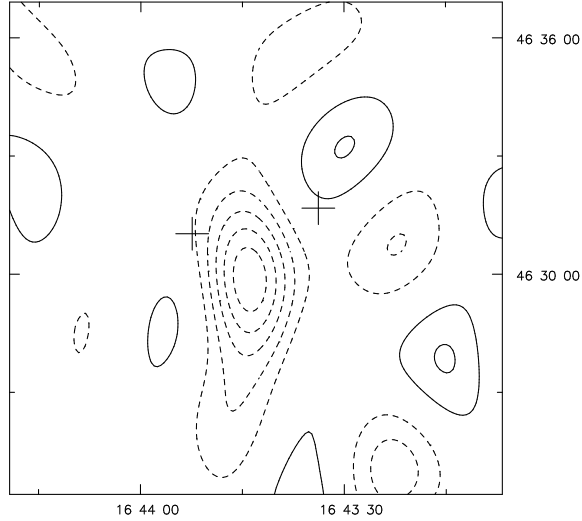,width=7cm,angle=-90}}
\caption{CMB decrement observed by the Ryle Telescope towards
PC1643+4631 A \& B (Jones et al. 1997).
The peak flux density is $-380\pm64\ \mu{\rm Jy\ beam^{-1}}$
while the noise level is $38\ \mu{\rm Jy\ beam^{-1}}$.
The two crosses indicate the positions of quasars A \& B.}
\label{fig:ryle}
\end{figure}

Recent observations carried out with the Ryle Telescope at 15 GHz
revealed an intriguing decrement in the CMB towards the pair of
quasars PC1643+4631 A \& B at red-shift $z \sim 3.8$ and
separated by 198 arc-seconds (Jones et al. 1997).
Fig.\ref{fig:ryle} shows the CMB decrement together with the quasar
positions. If this decrement is due to a S-Z effect, this is
indicative of a rich intervening galaxy cluster of total mass
$M \sim 10^{15} \ {\rm M_{\odot}}$. Since the quasars' spectra,
as observed with the William Herschel Telescope (WHT), are
virtually identical, Saunders et al. (1997) suggest that
such a massive cluster can account for the spectral similarities if
we assume that the quasar pair is lensed.
The remarkable aspect of the problem is that no X-ray cluster
is visible in this direction. Jones et al. (1997) argue that
the absence of a detection with the X-ray Roentgen Satellite
(ROSAT) indicates a redshift lower limit of $z\geq 1$ for the
expected cluster.
Saunders et al. (1997) report colour imaging made with
the WHT (R band) and with the UK Infrared Telescope (UKIRT)
(J and K bands). The absence of cluster in these observations
raises the limit up to $z \geq 1.5$. Recent deep
observations (work in progress) have been made with the WHT
(U,G,V,R and I bands). Again, no visible
clustering is observed, although there exist many objects in the
field with expected redshift of $z\sim 1-3$.
Moreover, the PC1643+4631 case is not isolated.
Indeed, a similar situation has also been reported from observations
of the CMB with the Very Large Array (VLA) (Richards et al. 1997).
The term ``dark cluster'' seems to have become increasingly popular
for such objects (e.g. Hattori 1997 et al., Jimenez et al. 1997,
Hawkins et al. 1997).\\

\section{Model}
\label{model}
Early attempts to estimate the Rees-Sciama effect induced by a
collapsing cluster of galaxies were based on the ``Swiss Cheese''
(SC) model (e.g. Rees \& Sciama 1968, Nottale 1982
\& 1984) whereas more recent attempts have used the continuous
Tolman-Bondi solution (e.g. Panek 1992). We
present here an improved model which, for the first time, treats
such a problem exactly.
%
%\begin{figure}
%\centerline{\epsfig{file=model.eps,width=12cm}}
%\caption{Fluid initial conditions for our model. Two parameters
%control the perturbation: its size $r_0$ and its rate of growth $a$.
%}
%\label{fig:model}
%\end{figure}
%
We assume that the cluster is in a state of pure infall,
at the centre of the reference frame.
Both the collapsing cluster and the expanding universe are
treated as a unique pressure-less fluid, the evolution of which
is determined by a set of analytical equations
(see Lasenby et al. 1997).
The fluid density and velocity field distributions
are continuous and realistic (See Dabrowski et al. 1997
and Fig.\ref{fig:distribution})
which is a major improvement over the SC models.
Indeed for the SC case, an artificial vacuum region is needed
to isolate the cluster from the universe which implies an
unrealistic cluster density profile as well as discontinuities
in the density and velocity field distributions.
Another advantage with regard to the SC
models is that our perturbation is initially defined by only
two free parameters rather than three in the SC case.
Initially, the density of our fluid is uniform and only the velocity
field is perturbed in order to form a realistic cluster.
In a given cosmology this perturbation is controlled by a size
parameter and a rate of growth parameter.
In addition to the evolution of the fluid we perform the exact
calculation of photon trajectories and redshifts
(Dabrowski et al. 1997) so that effects such as Rees-Sciama or
gravitational lensing can be fully appreciated.

\section{Results}
\label{results}
We now apply our model to the PC1643+4631 case in order to
estimate the typical properties of clusters which may explain
the observations of Jones et al. (1997).
Our aim is to account for both the CMB decrement and the
lensing of the quasar pair. We assume that the lensing effect
will be correctly taken into account if the Einstein
ring radius is 100 arc-seconds.
Throughout this paper, we assume the following cosmology:
$\Omega (t)=1$, $H_0=50\  {\rm km\  s^{-1}\  Mpc^{-1}}$.
\begin{figure}
\centerline{\epsfig{file=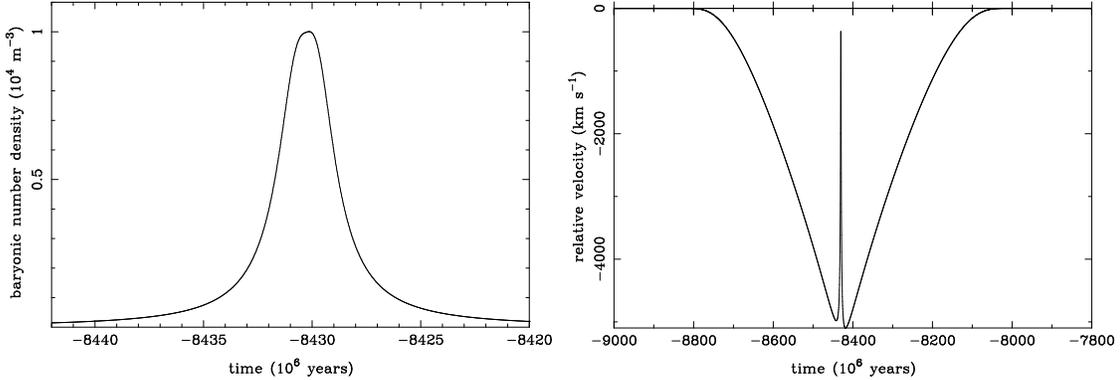,width=14.8cm}}
\caption{(left) Baryonic number density as a function of
proper time $t$ as experienced by a photon traversing the centre
of the cluster. A time $t=0$ would correspond to the present
epoch. (right) As for the left case but for the fluid velocity
relative to the Hubble flow.
}
\label{fig:distribution}
\end{figure}
In order to obtain quantitative results we need to constrain our
model parameters. We assume that the cluster lies at a redshift
$z=1$ and that $z=3.8$ for both quasars A \& B. We also consider
a very rich cluster and fix the maximum baryonic number density
experienced by an observed photon to $10^4\  {\rm m^{-3}}$ and
the total gravitational mass over baryonic mass ratio
is taken to be 10. In order to obtain an Einstein ring radius
of 100 arc-seconds we find that the core radius $R_c$ has to
be $0.45\ {\rm Mpc}$. $R_c$ is defined as the radius at which
the cluster energy falls to one-half its maximum value.
We find that the total mass contained within a sphere of 2 Mpc
is $7.8\times 10^{15}\ {\rm M_{\odot}}$ and
$1.9\times 10^{16}\ {\rm M_{\odot}}$ within a sphere of 4 Mpc.
The density and fluid velocity experienced by a photon
passing through the centre of the cluster are shown in
Fig.\ref{fig:distribution}. We find that the density profile can
be fitted accurately (Dabrowski et al. 1997) by a standard
King profile model known to represent realistically observed
clusters (Rood et al. 1972 \& Sarazin 1988).\\

We now investigate the effect of such a cluster on the CMB.
Firstly we compute the temperature $T_e$ required to produce
a total S-Z flux of $-380\ \mu{\rm Jy}$ on the Ryle Telescope.
We find $T_e=1.78\times 10^{7}\ {\rm K}$ and that the
corresponding S-Z temperature decrement is $\Delta T_{\rm SZ}
= 458\ \mu{\rm K}$. Since our model allows photon path and
redshift calculation we are able to estimate the Rees-Sciama
effect and we find $\Delta T_{\rm RS}=178\ \mu{\rm K}$. We
then conclude that the Rees Sciama effect is not negligible
and might contribute significantly to the total observed CMB
decrement.

%-----------------------------------------------------------------
%  end of text
%-----------------------------------------------------------------
%-----------------------------------------------------------------
\end{document}